\documentclass{ws-jai}
\usepackage[flushleft]{threeparttable}
\usepackage{xspace, hyperref}
\usepackage{subcaption}

\usepackage{color}

\begin{document}

\def \alb {ALBATROS\xspace}
\def \drone {PteroSoar\xspace}
\def \mars {MARS\xspace}

\catchline{}{}{}{}{} % Publisher's Area please ignore

\markboth{Lawrence Herman}{Drone-Based Antenna Beam Calibration in the High Arctic}

\title{Drone-Based Antenna Beam Calibration in the High Arctic}

\author{Lawrence Herman$^{1,5}$, Christopher Barbarie$^{2,1}$, Will Tyndall$^{1,5,\dagger}$, Mohan Agrawal$^{1,5}$, Vlad Calinescu$^{1}$, Simon Chen$^{1}$, H.~Cynthia Chiang$^{1,5}$, Aman Chokshi$^{1,5}$, Cherie K.\ Day$^{1}$, Eamon Egan$^{1}$, Stephen Fay$^{1}$, Kit Gerodias$^{1}$, Maya Goss$^{1}$, Michael H\'etu$^{1}$, Daniel C.\ Jacobs$^{3}$, Marc-Olivier R.\ Lalonde$^{1,3}$, Francis McGee$^{1,5}$, Lo\"ic Miara$^{1}$, John Orlowski-Scherer$^{1,4}$, and Jonathan Sievers$^{1,5}$}

\address{
$^{1}$Department of Physics, McGill University, Montr\'eal, QC H3A 2T8, Canada)\\
$^{2}$Department of Electrical, Computer \& Energy Engineering, University of Colorado Boulder, Boulder, CO 80303, USA\\
$^{3}$School of Earth and Space Exploration, Arizona State University, Tempe, AZ 85287, USA\\
$^{4}$Department of Physics and Astronomy, University of Pennsylvania, Philadelphia, PA 19104, USA\\
$^{5}$Trottier Space Institute, McGill University, Montréal, Québec H3A 2A7, Canada\\
}

\maketitle

\corres{$^\dagger$Corresponding author: william.tyndall@mcgill.ca}

\begin{history}
\received{(to be inserted by publisher)};
\revised{(to be inserted by publisher)};
\accepted{(to be inserted by publisher)};
\end{history}

%%%%%%%%%%%%%%%%%%%%%%%%%%%%%%%%%%%%%%%%%%%%%%%%%%%%%%%
%%                     Abstract                      %%
%%%%%%%%%%%%%%%%%%%%%%%%%%%%%%%%%%%%%%%%%%%%%%%%%%%%%%%

\begin{abstract}

\vspace{3mm}
Precision calibration is a critical requirement for future ultra-low-frequency observations of the early universe. The Array of Long Baseline Antennas for Taking Radio Observations from the Seventy-Ninth Parallel (ALBATROS), a radio interferometer located in the Canadian high Arctic, is designed to map Galactic foreground emission as a pathfinder for these future experiments. Accurate antenna beam characterization at these frequencies is therefore essential, yet remains uniquely challenging. We present PteroSoar, a custom-built drone platform equipped with a calibrated radio-frequency transmitter that enables controlled, in-situ measurements of low-frequency antenna beams. As an initial demonstration, we produce a two-dimensional beam map of an ALBATROS antenna at 50~MHz to a precision of approximately $10\%$ near zenith. We identify the dominant sources of systematic uncertainty, including timing imprecision, and outline hardware and software improvements that are expected to reduce beam measurement uncertainties to below $5\%$. This target is sub-dominant to the $\sim20\%$ amplitude variability introduced by ionospheric scintillation at these frequencies, providing a practical pathway toward precision beam calibration for ALBATROS and other ultra-low-frequency radio experiments.

\end{abstract}


%%%%%%%%%%%%%%%%%%%%%%%%%%%%%%%%%%%%%%%%%%%%%%%%%%%%%%%
%%                   Introduction                    %%
%%%%%%%%%%%%%%%%%%%%%%%%%%%%%%%%%%%%%%%%%%%%%%%%%%%%%%%

\section{Scientific Background}
\label{sec:Intro}

The observational history of the Universe is missing a key chapter: the Dark Ages, Cosmic Dawn, and the Epoch of Reionization, which bridge the gap between the Cosmic Microwave Background and the mature, structured Universe observed today. During these epochs, the first stars, galaxies, and black holes formed, illuminating and gradually reionizing the neutral intergalactic medium \citep{FURLANETTO_OH_BRIGGS_2006, PRITCHARD_LOEB_2012}. Observing these epochs remains one of the foremost challenges in modern observational cosmology, promising to complete our empirical picture of cosmic evolution.

The redshifted 21-cm transition of neutral hydrogen provides the primary observational probe of these epochs, motivating a new generation of low-frequency radio experiments. These include global signal experiments such as EDGES \citep{EDGES_Cappallo_2025PASP..137l5002C}, SARAS \citep{SARAS_Nambissan_2021ExA....51..193T}, and MIST \citep{MIST_Monsalve_2024MNRAS.530.4125M}, together with interferometric arrays including MWA \citep{MWA_Wayth_2018PASA...35...33W}, HERA \citep{HERA_DeBoer_2017PASP..129d5001D}, LOFAR \citep{LOFAR_vanHaarlem_2013A&A...556A...2V}, and the forthcoming SKA-Low \citep{SKA_LOW_Mellema_2013ExA....36..235M}. While current facilities primarily target the Epoch of Reionization above $\sim100$ MHz, extending observations toward the lower frequencies corresponding to Cosmic Dawn and the Dark Ages remains a major frontier in experimental cosmology.

Pushing 21-cm observations to progressively lower frequencies---and hence higher redshifts---becomes increasingly challenging from the ground. The Earth’s ionosphere introduces increasingly severe propagation effects, including refraction, scintillation, high spatial and temporal variability, while ultimately imposing a plasma-frequency cutoff. Although facilities such as OVRO-LWA \citep{OVRO_LWA_EASTWOOD_2018AJ....156...32E} and NenuFAR \citep{NenuFAR_Zarka_2015icat.conf...10Z} have extended ground-based observations into the tens of megahertz, the exceptionally low plasma frequencies of $\lesssim 10$ MHz remain under-explored. The state-of-the-art at the lowest observable frequencies dates back to the 1960s when Reber, Ellis, and others caught brief glimpses of the Galaxy around 5 MHz from Tasmania \cite{ellis1982galactic}. Terrestrial access to cut-off frequencies much lower than 10 MHz is provided by the polar ionosphere, where the plasma frequency can drop to 2-3 MHz during periods of quiet solar activity during polar night \citep{BJOLAND_POLAR_IONO_2021, GRUNEY_THEMENS_IONO_2025}. The Array of Long Baseline Antennas for Taking Radio Observations from Seventy-Ninth Parallel \citep[ALBATROS;][]{2020JAI.....950019C}, located on Umingmat Nunaat (Axel Heiberg Island, Nunavut; 79$^\circ$N), was established to exploit this unique environment. One of the primary science goals of ALBATROS is to image the northern sky between $\sim$4--40~MHz with few-arcminute resolution to address the sparsity of high-resolution sky maps in this frequency band. By accurately characterizing the Galactic emission, the ALBATROS maps will not only act a pathfinder for future Dark Ages experiments, but also potentially open a new observational window onto the ultra-low-frequency Universe.

Producing images of the sky with ALBATROS requires precisely characterizing the instrument.  We present \drone, a new drone-based system that is designed to fly in the high Arctic and characterize the \alb on-sky spatial response, or beam pattern.  Figure~\ref{fig:albatros} illustrates \alb and \drone operating on Umingmat Nunaat.  This paper is organized as follows: \S\ref{sec:drone_intro} gives an introduction to drone-based beam measurements, \S\ref{sec:iono_precision} sets a benchmark for the required precision of \alb beam measurements, \S\ref{sec:methodology} describes the \drone design and operations, \S\ref{sec:processing} discusses the data processing and analysis pipeline, \S\ref{sec:results} presents \alb beam mapping results from an initial demonstration measurement performed during the 2023 Arctic summer, and \S\ref{sec:conclusions} discusses future improvements and \alb beam mapping plans.

\section{Drone-based Beam Measurements}\label{sec:drone_intro}

Cosmology experiments at radiofrequencies depend on measurements of the beam pattern for calibration. The level of precision, however, depends on the science case. For example, estimates suggest that foreground subtraction for Epoch of Reionization experiments will require precision beam measurements to better than 1\% uncertainty \cite{shaver1999, oh2003, liu2011, Shaw_polmodes, CC_drone_mapping}. Radio telescopes and interferometer arrays are often calibrated in this way by receiving emissions from celestial sources that have been extremely well-studied\cite{standardsources, Nunhokee_HERA_beam_2020, Berger_CHIME_2016, Pober_DRIFT_SCAN_2012, Bowman_Beam_2007}. Experiments that outfit their antennas with active drive systems have the capacity to steer their elements in two spatial dimensions across these bright celestial sources, directly assessing the relative antenna gain by comparing the received power measured in different portions of the beam. However, this technique is not an option for experiments with stationary antennas---a common design choice for transit interferometers like \alb. Such experiments have successfully used astronomical and satellite source transits to obtain one-dimensional beam profiles (measured each day in the East-West direction as the earth rotates), which can be combined into a two-dimensional map by aggregating repeated measurements as the sources altitude changes each day~\cite{2022ApJ...932..100A, 10.1093/mnras/stab156, 2018PASA...35...45L, 2015RaSc...50..614N}. This measurement process necessarily takes place on long time scales, and requires careful treatment of instrument systematics. 

An alternative method of beam calibrating these stationary instruments is to probe the instrument field of view with a drone or UAV (Uncrewed Aerial Vehicle) carrying a transmitting radiofrequency source. Drone-based calibration systems for radio astronomy have been demonstrated in the literature numerous times, using a variety of designs and instrumental approaches~\cite{ Virone2014, Pupillo2015, CC_drone_mapping, 2016SPIE.9906E..3VB, 2017PASP..129c5002J, 9560255, 2022JAI....1150016B, ToneBeamcals, CHIMETyndall}. In addition to the increased mapping speed, several other advantages emerge from this method. First, because drones have negligible angular sizes at relevant distances from the telescope and thus resemble point sources at radio frequencies, measurements precisely reproduce the instrument beam. Second, the relative motion of drones can be controlled with a flightplan designed to raster across the instrument field of view, measuring two-dimensional beams with arbitrarily fine angular resolution. Third, artificial sources can be designed with precision control over the amplitude and frequency structure of the broadcast signal. A final advantage is that antenna performance is being characterized on site, within the configuration of the larger array and surrounding environment \cite {2017PASP..129c5002J, MauermayerNFdrone}. 

After a decade of technical innovations, commercial drone platforms have significantly improved, increasing the feasibility of using drones as a precision instrument for calibrating radio telescopes. One method of improving the precision of the drone is augmenting on-board GPS systems with additional equipment, including real-time kinematic (RTK) systems, differential GPS systems, or both \cite{GarciaFernandez2017,GarciaFernandez2019,CulottaLopez2021,Zhang2021}. 
In either configuration, additional information from a secondary GPS receiver enhances the accuracy of the GPS receivers on the UAV, reducing positioning uncertainty; in some cases, as low as $\sigma_{\vec{x}}\sim1$\,cm. Many such systems (like the DJI D-RTK) are provided by the same manufacturer as the UAV, and communicate directly with the UAV during flight operations\cite{oconnor2020baryon,ToneBeamcals,CHIMETyndall,Zhang2021}). This communication layer provides real-time position corrections to the drone at the level of the flight controller, improving stability and positional accuracy, while reducing the magnitude of deviations from the intended flightplan. With high precision positional accuracy, autopilot software, and carefully designed flightplans, UAV calibration can directly measure the far-field beam pattern of telescopes and arrays, where the beam properties are fully evolved and the beam profiles are identical to the sky beam used in cosmological observations. 

\section{Precision Criteria and Ionospheric Scintillation}\label{sec:iono_precision}

The primary limitation to ALBATROS' long-term goal of imaging the the sky across the HF band ($\sim$4--40~MHz) is ionospheric scintillation, whose effects become increasingly severe at lower frequencies, particularly in the polar region \cite{aarons1971global}. For the initial map-making efforts, we require uncertainties in the measured beam to remain sub-dominant to those arising from ionospheric scintillation. As the present work characterizes only the antenna power beam, we compare our measurement uncertainties with the expected level of ionospheric amplitude scintillation in order to develop a precision criterion that can guide our beam characterization efforts.

The strength of ionospheric amplitude scintillation is conventionally quantified by the scintillation index, $\mathrm{S}_4$, defined as the normalized variance of fluctuation of the received signal intensity $I$,
\begin{equation}
    \mathrm{S}_4 = \sqrt{\frac{\langle I ^2 \rangle - \langle I \rangle^2}{\langle  I\rangle^2}}.
\end{equation}
Earlier studies occasionally adopted the empirical scintillation index (S.I.), defined as
\begin{equation}
    \mathrm{S.I.} = \frac{R_{max} - R_{min}}{R_{max} + R_{min}},
\end{equation}
where $R$ could either be the intensity or the amplitude of the recorded signal.

Results on ionospheric scintillation in the polar regions near the upper-end of ALBATROS science band $\sim$40~MHz are sparse in literature. An estimate of typical scintillation levels at high latitudes ($> 65^\circ$) can be obtained using historic records of amplitude scintillation data obtained using observations of satellites with carrier-wave radio beacons in the 1960s. A year of observations in 1965-1966, corresponding to the minimum activity phase of the 20th solar cycle, made from Spitsbergen (78$^\circ$N) report an amplitude S.I. of 0.2 for quiet geomagnetic conditions (Kp $<$ 3), and 0.4 for active geomagnetic conditions (Kp $\ge$ 3). Observations of the same satellite beacon made by \citealp[]{joint1968latitude} in winter 1965 from Narssarssuaq (71$^\circ$ N) report power S.I. of about 0.65. In the same study, a year of observations of a different satellite beacon (Transit-4A) at 54 MHz made from  Kiruna (65$^\circ$ N) between 1961--1962 suggest a power S.I. of 0.33. While there's no straightforward formula for converting the historical amplitude/power S.I. metric to the modern $\mathrm{S}_4$ metric, empirical relationship suggests that the distribution of received signal's amplitude is well-approximated by the Nakagami-m distribution \cite{fremouw1980statistics, bischoff1969note, whitney1972estimation}. Their results indicate that the typical $\mathrm{S}_4$ for the S.I. reported by aforementioned results lies in the range 0.2--0.3. We take an $\mathrm{S}_4$ of 0.2 to be suggestive of quiet ionospheric conditions during solar cycle minimum in the polar cap. During the peak of 20th solar cycle (1968-1969) \citealp[]{Singleton1970-ot} reported typical scintillation index of 0.5 from Thule (72$^\circ$ N), with their definition of scintillation index equivalent to $\mathrm{S}_4$\cite{singleton1970effect, briggs1963variation}. Additionally, analysis of Explorer-22 observations taken aboard the HMNZS Endeavour from the southern polar cap (during its voyages between New Zealand and Antarctica) between 1966--1968, also shows a high $\mathrm{S}_4$ of about 0.75 albeit with a very high scatter for latitudes between 70$^\circ$-80$^\circ$S. 

Despite the variety of scintillation metrics adopted throughout the literature, these measurements present a consistent picture, with typical $\mathrm{S}_4$ of 0.2--0.3 during solar minimum, and increasing to more than 0.5 near solar maximum. The natural variability introduced by the ionosphere thus provides a practical benchmark for beam calibration. To ensure that beam uncertainties remain sub-dominant to $\mathrm{S}_4$, we adopt an initial design target of 5\% uncertainty in the measured beam response. For an in-depth error analysis see Appendix \ref{subsec:ionosphere_appendix}.

%%%%%%%%%%%%%%%%%%%%%%%%%%%%%%%%%%%%%%%%%%%%%%%%%%%%%%%
%%                   Methodology                     %%
%%%%%%%%%%%%%%%%%%%%%%%%%%%%%%%%%%%%%%%%%%%%%%%%%%%%%%%

\section{The \drone System}
\label{sec:methodology}

Beam mapping operations require long flight times with payloads that are typically $\sim$1~kg. Commercially available drones that meet these design criteria are expensive, typically around \$15,000. This high cost presents a significant risk for remote Arctic operations, where flyaways, crashes, or hardware failures can result in complete drone loss. 

\begin{figure*}[t]
  \centering
  % use slightly smaller combined width to allow for small gap
  \begin{subfigure}[t]{0.545\textwidth}
    \centering
    \includegraphics[height=0.33\textheight,trim=0 0 0 0,clip]{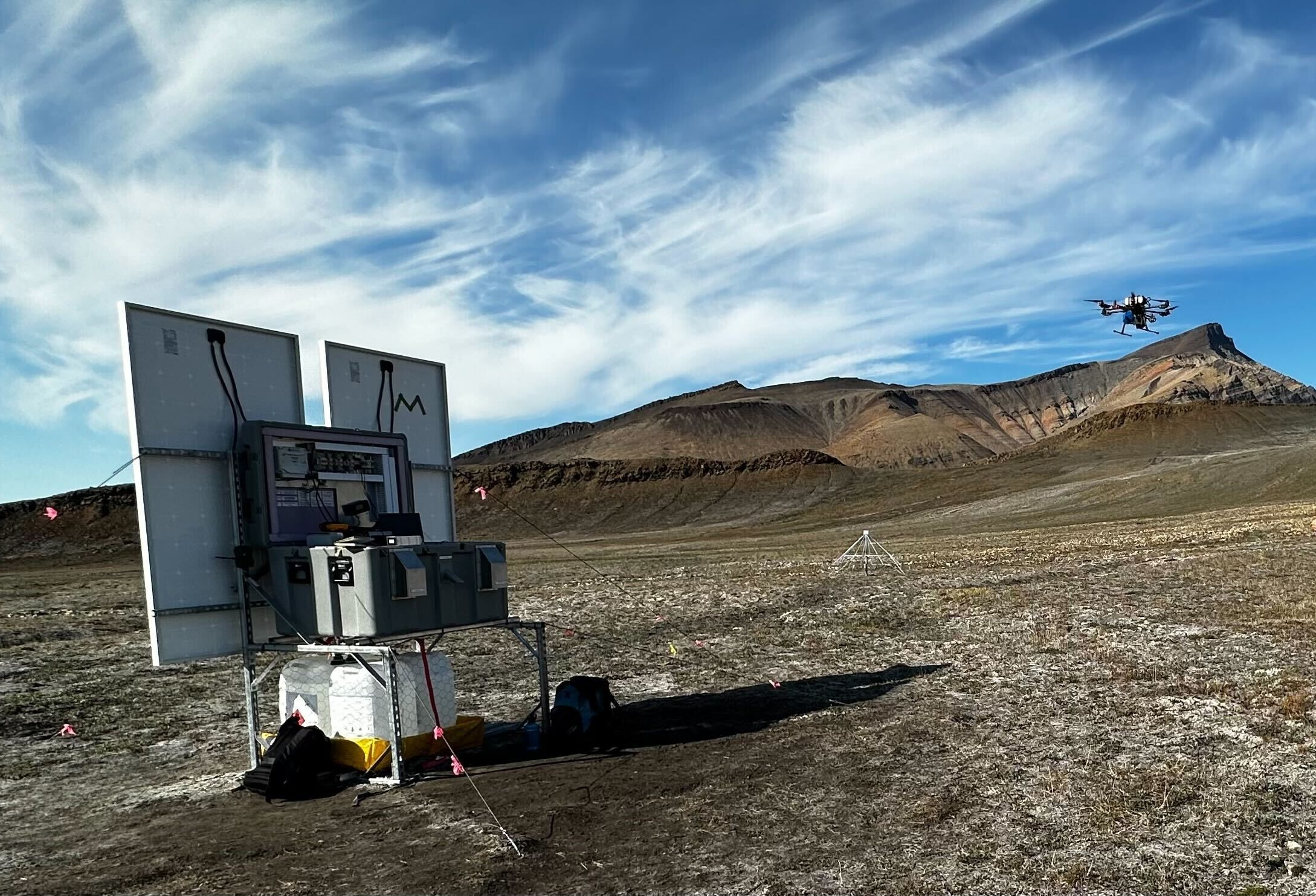}
    \caption{}
    \label{fig:albatros_a}
  \end{subfigure}%
  \hspace{0.015\textwidth}% small gap (~1.5% of width)
  \begin{subfigure}[t]{0.44\textwidth}
    \centering
    \includegraphics[height=0.33\textheight,trim=0 0 0 0,clip]{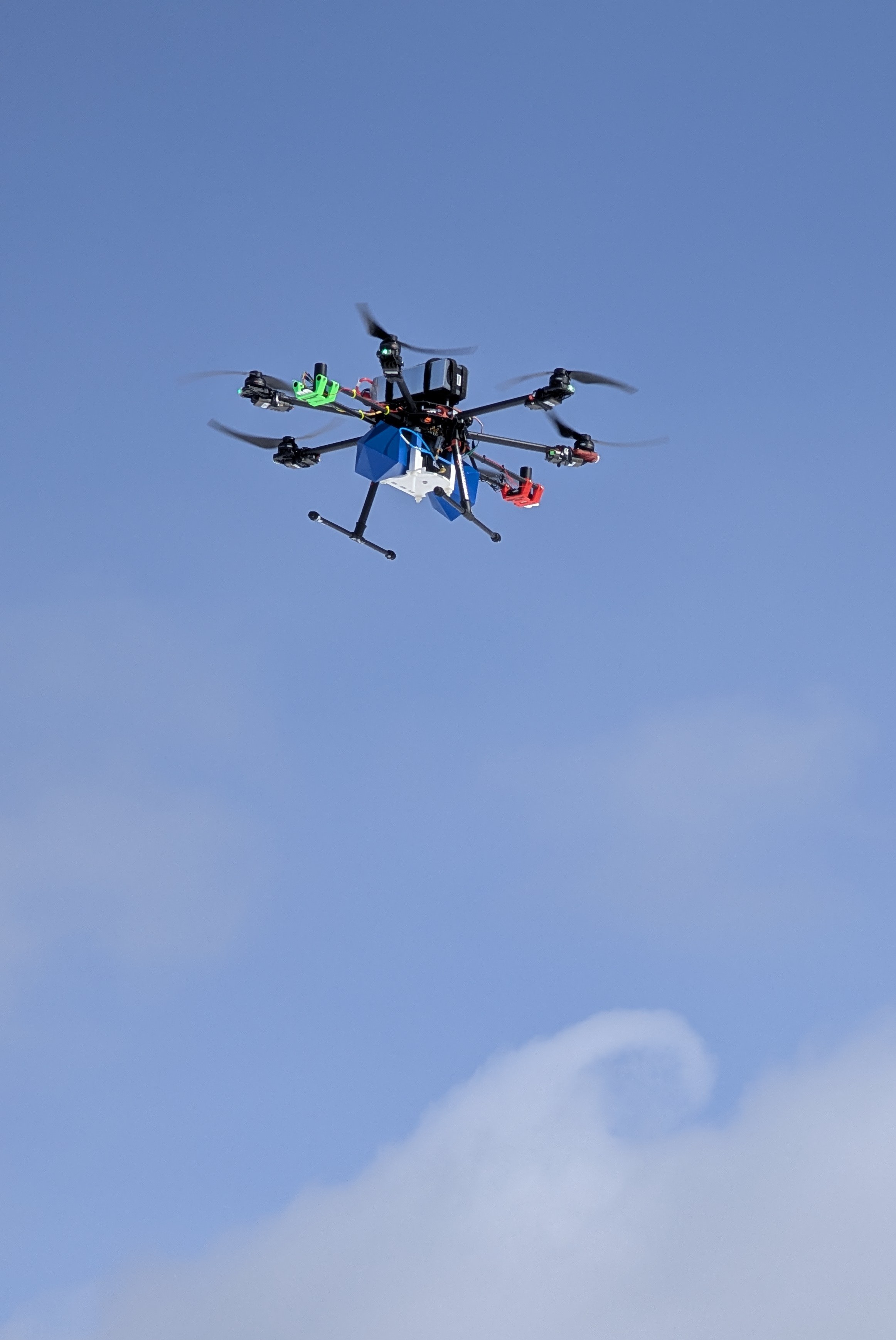}
    \caption{}
    \label{fig:albatros_b}
  \end{subfigure}

  \caption{(a) \drone in flight over an \alb station located on Axel Heiberg Island at \mars, between the receiving antenna (background) located 30~m from the readout and power systems (foreground). (b) Close-up view of the \drone carrying the calibration source.}
  \label{fig:albatros}
\end{figure*}

A custom design offers several advantages including subsystem modularity, tailored control of design, and dramatically reduced cost. Additionally the ability for quick and cheap field-repair is crucial for remote deployments. By comparison, many commercial systems are not user-serviceable, and any subsystem failures at a remote deployment site would halt all drone operations. The \drone system employs a custom-built hexacopter (see Figure \ref{fig:albatros}), designed for precision beam mapping of low-frequency radio antennas, including ALBATROS\cite{2020JAI.....950019C}, MIST\cite{MIST_Monsalve_2024MNRAS.530.4125M} and HIRAX\cite{hirax}.

%%%%%%%%%%%%%%%%%%%%%%%%%%%%%%%%%%%%%%%%%%%%%%%%%%%%%%%
%%                    The Drone                      %%
%%%%%%%%%%%%%%%%%%%%%%%%%%%%%%%%%%%%%%%%%%%%%%%%%%%%%%%

\subsection{The \drone Drone}
\label{subsubsec:drone}

\begin{figure}[bh]
\centerline{\includegraphics[width=0.9\textwidth]{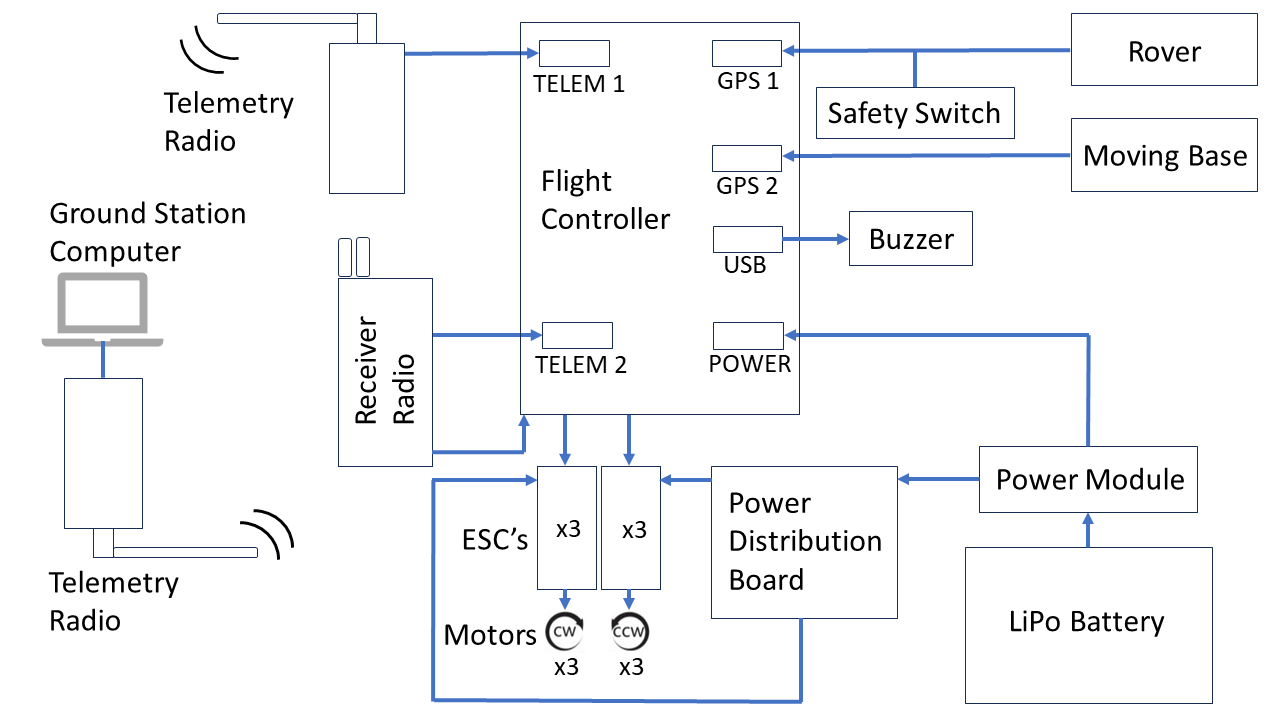}}
\caption{Block diagram of drone components. The drone is controlled by a Pixhawk Cube Orange flight controller which synthesizes information from the dual-GPS system (GPS1/GPS2) and a receiver radio (TELEM2) to command the electronic speed controllers (ESCs) and motors. The motors drive three pairs of counter-rotating (clockwise (CW) and counter-clockwise (CCW)) propellers for balanced torque. The flight controller is also equipped with a telemetry radio (TELEM1) which relays positioning data to a ground station computer. The entire system is powered by a lithium polymer battery.}
\label{fig:components}
\end{figure}

The \drone drone design (Figure~\ref{fig:components}) is based largely on the External Calibrator for Hydrogen Observatories (ECHO)~\cite{2017PASP..129c5002J} and is constructed from off-the-shelf components. The drone utilizes a Tarot Iron Man 680 carbon fiber frame for a high strength-to-weight ratio with overall dimensions 1015~mm~$\times$ 1015~mm~$\times$ 330~mm. A block diagram of the drone circuit components is presented in Figure~\ref{fig:components}. Power is supplied by a single Tattu 6S lithium-polymer 17000~mAh battery. Each battery provides approximately 30~minutes of flight time at an altitude of $\sim70$~m above ground level (AGL). The motors, propellers, and electronic speed controllers (ESCs) are from the DJI E800 kit. Each of the six motors produces 2100~g of thrust. The flight platform (drone \& batteries) is approximately 4~kg.  To maintain a 2:1 thrust-to-weight ratio, the maximum payload the drone can carry is approximately 2.8~kg. 

A Pixhawk Cube Orange flight controller, mounted on the drone frame central platform, controls the drone's attitude and motion. The drone can accept either a pre-programmed flight path from the flight controller or pilot navigation. The flight controller also contains a barometric sensor for altitude estimation and three orthogonal integrated accelerometers and gyroscopes, for attitude and motion sensing. Real-time flight telemetry data is streamed to a laptop base station via a 915-MHz, 100-mW radio link. Position estimation and heading input are provided via a dual-GPS system using Holybro F9P Helical GNSS modules. Open-source software from QGroundControl\footnote{\url{https://qgroundcontrol.com/}} is used for flight planning, autopilot control of the drone, and configuring the drone’s various software parameters. A FrSky Taranis X9D Plus SE 2019 handheld controller provides manual control of the drone via a 2.4-GHz, 32-mW radio link. While manually flying the drone, roll and pitch are controlled via acceleration command and yaw via rate command.

\subsection{The dual-GPS and Real-Time Kinematic systems}
\label{subsubsec:dgpsrtk}

\drone incorporates a Real-Time Kinematic (RTK) system consisting of a ground base station and a dual-GPS pair of uBlox F9P GPS modules\footnote{\url{https://www.u-blox.com/en/product/zed-f9p-module}} onboard the drone that are separated by a fixed 66~cm baseline. The RTK module enables us to determine the drone position to sub-centimeter accuracy at $\sim$10~Hz, comparable to existing drone beam mapping experiments\cite{2017PASP..129c5002J,CHIMETyndall,ToneBeamcals}. 

While most drones determine their heading with magnetometer readings---which are not reliable in the high Arctic---we do so using our dual-GPS system. Near the Magnetic North Pole in the high Arctic, magnetic field lines are nearly vertical (at \mars, the magnetic field inclination is $\sim88^{\circ}$,) reducing magnetometer sensitivity\cite{magnetic}. The differential dual-GPS configuration onboard \drone reliably achieves sub-degree heading accuracy. Crucially, this translates to an equivalent precision in the polarization direction of the transmitting antenna which is securely mounted to the drone chassis.

\subsection{QGroundControl: Mission Planning and Ground Control Software}
\label{subsubsec:qgc}

The capabilities of \drone are greatly enhanced by QGroundControl, an open source software program for mission planning and flight control. During mission planning, a flight path can be designed to raster over an instrument for optimal sampling of a telescope beam (see Section\ref{subsubsec:flightplanning}). QGroundControl is also capable of autopiloting the drone during flight, progressing through a series of waypoints without input from a pilot. While a pilot can manually control the drone via remote control, automation ensures steady flight at a fixed speed during beam mapping, reducing uncertainty in the position and orientation of the source of the calibration signal.

\subsection{Calibrator Payload}
\label{subsec:payload}

\begin{figure}
\centerline{\includegraphics[width=3.5in]{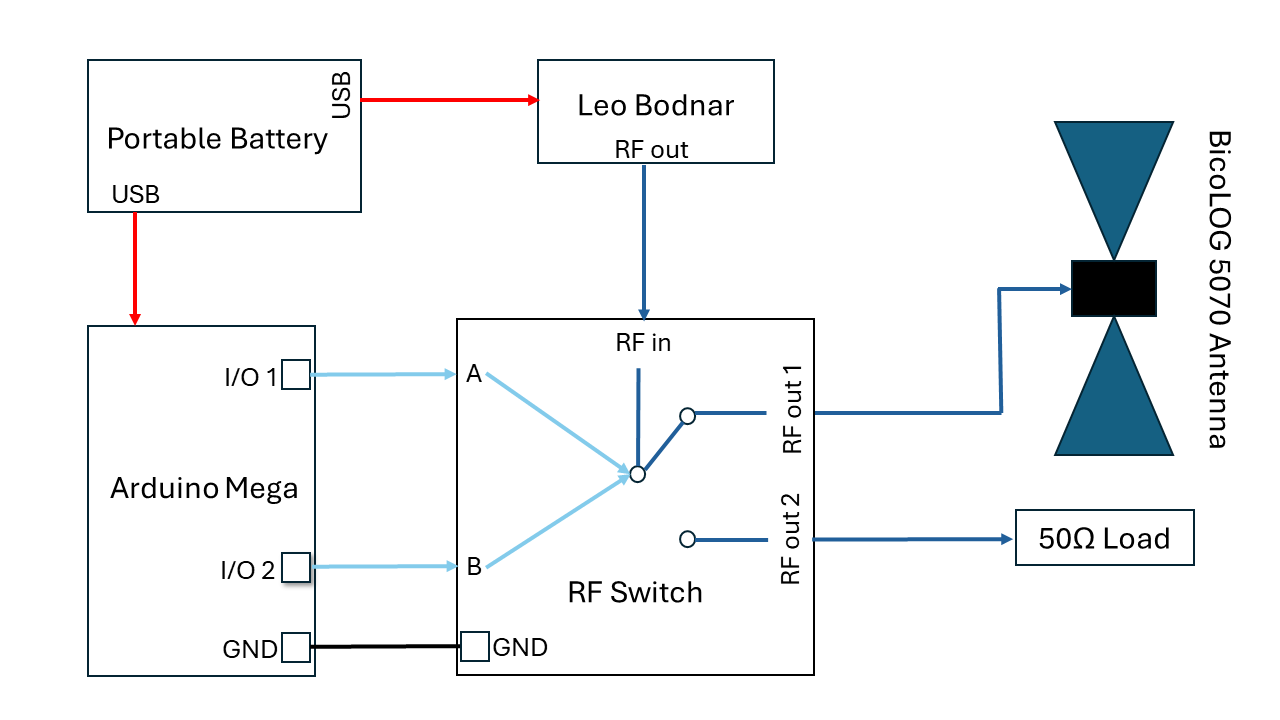}}
\caption{Block diagram of the transmitter payload. A 5-V portable battery powers all active components (red arrows). A Leo Bodnar mini precision GPS reference clock produces a calibration signal (dark blue arrows) which is chopped by an RF switch. An Arduino Mega controls the RF switch via logic circuitry (cyan arrows), which alternates between a 50~$\Omega$ terminator and the BicoLOG 5070 antenna.}
\label{fig:payload}
\end{figure}

The calibrator payload of the \drone system transmits a precisely tuned and chopped signal, enabling the mapping of antenna beams. A block diagram of the \drone calibrator payload is shown in Figure~\ref{fig:payload}. 

The transmitted radio signal is generated by a Leo Bodnar\footnote{\url{https://www.leobodnar.com/shop/index.php?main_page=product_info&products_id=301}} mini precision GPS reference clock which produces a square-wave tunable from 400~Hz to 810~MHz. The output power is approximately 10~dBm, sufficient for \alb beam measurements at our test frequency of 50~MHz without additional amplification. An Arduino Mega and a HMC545 RF switch chops the calibration signal at 10~Hz against a 50-$\Omega$ terminator to facilitate the separation of the calibration and background signals in post-processing (see Section \ref{sec:processing}). The 10~Hz chopping frequency is tunable, and is chosen to be compatible with the flight speed and data acquisition rates (further details in~\S\ref{subsubsec:flightplanning}). 

The calibration signal is transmitted by an Aaronia BicoLOG 5070 antenna\footnote{\url{https://aaronia.com/en/produkte/antennas/bicolog}}, which is commercially available and well characterized. BicoLOG antennas are commonly used for drone beam mapping because they transmit a single linear polarization with a broad beam that is spatially and spectrally smooth\cite{2017PASP..129c5002J,CHIMETyndall,ToneBeamcals}. By aligning the antenna with the nose of the drone, the polarization direction of the calibration signal is locked to the drone heading.  We have performed numerical electromagnetic simulations of the BicoLOG antenna and the drone chassis using FEKO\footnote{\url{https://altair.com/altair-hyperworks}}. While the BicoLOG antenna was ideal for prototyping \drone, calibrating \alb at lower frequencies (5-30~MHz) will require a different transmitting antenna with similar characteristics (see Section \ref{sec:conclusions}).

A custom 3D-printed mounting system houses the payload components and secures them to the drone frame. In the event of a hard landing, the 3D-printed mount is designed to break away from the frame to minimize damage to the drone and the payload. When assembled, the gross weight of the payload is approximately 1~kg.

%%%%%%%%%%%%%%%%%%%%%%%%%%%%%%%%%%%%%%%%%%%%%%%%%%%%%%%
%%                   Drone Flights                   %%
%%%%%%%%%%%%%%%%%%%%%%%%%%%%%%%%%%%%%%%%%%%%%%%%%%%%%%%

\subsection{Drone Flights}
\label{subsec:droneflights}

\subsubsection{Flight Planning}
\label{subsubsec:flightplanning}

To sample the beam of an Antenna Under Test (AUT), \drone flies automated grid patterns over the AUT field of view while transmitting a calibration signal. When designing a flightplan to map an AUT, any pattern that sufficiently covers the field of view of the instrument can be used, but conventional choices like Cartesian, polar, and spherical shell raster patterns are frequently used in the literature\cite{2017PASP..129c5002J,CHIMETyndall,ToneBeamcals}. 

In the following subsections, we will motivate the specific constraints and requirements we have satisfied when designing \drone, and describe the flight pattern we have used to measure the main beam of an \alb antenna: a Cartesian raster at a fixed altitude of 60~m AGL. There are many requirements that must be considered when designing an optimal sampling pattern. Far-field altitude, spatial coverage, and angular resolution are the highest priorities, but there are also technical limitations of any drone system that must be considered. For \drone, these include battery life ($\sim$30~minutes), the drone data timestamping frequency ($\sim$10~Hz), the calibration source pulse frequency ($\sim$10~Hz), and AUT time resolution (e.g. $\sim$8~ms for ALBATROS data).
 
\subsubsection{Flight Altitude, Spatial Coverage, and Angular Resolution}
\label{subsubsec:altitude}

When beam mapping, the far-field distance of the antenna is typically a lower limit on flight altitude. The far-field distance for the electrically short and resonant antenna cases are $2\lambda$ and $2D^2/\lambda$, respectively, where $\lambda$ is the wavelength and $D$ is the antenna diameter\cite{stutzman2012antenna, Balanis}. At the calibration signal frequency of 50~MHz ($\lambda\sim6$~m) the $D\sim$2~m diameter \alb antenna falls between these two cases. Thus the far-field distance of 12~m (from the electrically short case) is an upper limit for the far-field distance of an \alb antenna at 50~MHz, and any flights performed at an altitude exceeding 12~m will be within the far-field. Without prior authorization from Transport Canada, drone operations are limited to a maximum flight altitude of 122~m AGL to prevent airspace violations. Testing revealed that ascent to and descent from 122~m AGL depletes approximately 20\% of the drone battery, while a lower altitude helped conserve power to extend the flight time available for grid flights. Additionally, a cartesian grid will sample a larger solid angle above the AUT when performed at a lower altitude, increasing the angular spatial coverage of the resulting beam map. We thus choose a flight altitude of 60~m AGL to place the drone well in the far field, while limiting battery power consumption during the ascent and descent. 

In future measurement campaigns, we intend to measure the beam of a single \alb antenna within an 80$^\circ$~$\times$~80$^\circ$ region with 1$^\circ$ of angular resolution. At an altitude of 60~m AGL, the desired spatial coverage corresponds to a square grid spanning approximately 100~m~$\times$ 100~m, while the desired angular resolution corresponds to a $\sim$1~m linear separation between the rows of the raster pattern.

\subsubsection{Flight Speed and Map Statistics}
\label{subsubsec:spacing}

After determining the spatial coverage and angular resolution requirements of the beam map, the drone speed needed to complete the grid pattern can be estimated, though several technical limitations must be considered before selecting a flight speed. Because it is often convenient to finish a grid pattern in a single flight, the flight duration is often limited to the $\sim$30~minute lifetime of a single battery, though larger patterns can be completed by changing batteries and conducting multiple flights. Additionally, the data cadence of each system involved in the measurement must be considered to avoid undersampling. The chopping rate of the calibration signal (10~Hz) and the time resolution of the drone position data (10~Hz) also impact flight speed. To perform background subtraction, we must obtain a spectra where the drone is both on and off within each pixel of the final map. Therefore, if the drone were to exceed a speed of $(1~{\rm m})/(0.1~{\rm sec}) = 10~{\rm m/s}$, the speed would not permit background subtraction in each pixel, failing to meet the 1$^\circ$ resolution requirement. Conversely, for \alb, the readout system digitizes the antenna timestreams at 250~Msps and saves channelized, 4-bit quantized voltage data at 50~MHz every 16.384~$\mu$s. This data rate is much faster than both the transmitter chopping and the drone position data rate, so the \alb readout system does not constrain the drone flight speed. However, instruments with longer integration times between spectra may constrain flight speed. 

Although the drone can operate comfortably at speeds in excess of 15~m/s, several advantages are apparent when the drone flies at lower velocities: the drone is more stable in flight, requires fewer autopilot corrections, and has improved position accuracy. Because the drone spends longer in each map pixel, multiple background subtracted measurements can be acquired by the AUT and averaged during data processing, improving per-pixel uncertainty in the final map. Finally, because drones move horizontally by pulling themselves through the air, higher lateral speeds imply a more severe tilt in the pitch and roll directions. The variance in pitch and roll during the flight propagates directly to uncertainty in the orientation of the transmitting antenna (and thus the amplitude of the calibration signal), because it is mounted directly to the drone chassis. For these reasons, we set the flight speed to approximately 1.5~m/s while mapping \alb, which yields a high density of measurement points necessary for background subtraction (see Section \ref{sec:processing}) and ensures each pixel of the beam at degree-scale resolution is measured at least once, satisfying the angular resolution requirement. At this speed, a complete beam map of a single antenna at an altitude of 60~m AGL will require two fully charged batteries.

%%%%%%%%%%%%%%%%%%%%%%%%%%%%%%%%%%%%%%%%%%%%%%%%%%%%%%%
%%                 Data Processing                   %%
%%%%%%%%%%%%%%%%%%%%%%%%%%%%%%%%%%%%%%%%%%%%%%%%%%%%%%%

\section{Data Processing}
\label{sec:processing}

The crux of drone beam mapping is synchronizing the drone position and telescope visibility data, thereby creating a positive association between the calibration signal transmitted from the drone at a precisely known position and the power measured by the antenna under test. Combining the two data streams was possible using \texttt{beamcals}\footnote{\hyperlink{https://github.com/WrightLaboratory/newburghlabdrone/tree/master/beamcals}{https://github.com/WrightLaboratory/newburghlabdrone/}}\cite{ToneBeamcals,CHIMETyndall}---an open source python package. For this analysis, a new branch was developed to handle the high time resolution of the \alb single dish data. In this section, we will describe the operations performed by the data processing and analysis pipeline used to produce our results. 

\subsection{Drone Data}

After each \drone flight, the flight controller generates log files that contain sensor data recorded alongside high precision timestamps obtained from GPS lock. The position, orientation, and altitude of the drone are determined by ``fused'' sensor data, obtained by combining readings from the drone’s inertial measurement units, GPS modules, and barometers. Using functions from the \texttt{beamcals} module, these coordinates are transformed from latitude, longitude, height to local (referenced to the antenna) Cartesian and Polar coordinate systems.

\subsection{\alb Baseband Data}
Each independent \alb station contains a Smart Network ADC Processor (SNAP\cite{SNAPhickish}) which digitizes time-series baseband volatage data at a rate of $250~\text{Msamp/s}$, saving locally on hard drives. The Arctic site location and long baselines, spanning 178m--8.7km, prevent real-time correlation and thus require offline correlation. These baseband data from all 8 stations can be interferometrically combined to produce auto- and cross-correlation spectra across a frequency range from $0-125~\text{MHz}$ with a time resolution of $16~\mu \text{s}$.

The physical limitations (e.g., battery power, flight time) of the \drone system limit the geographic scale of our drone flights to a fairly small area---relative to the baseline separation of the \alb stations---such that only one antenna can be mapped at a time. Consequently, our calibration flights only require auto-correlation spectra from a single dish. Moreover, because the RF payload produces a calibration pulse at a single frequency ($50~\text{MHz}$) which is broadcast in a single linear polarization via the transmitting antenna, our maps from each flight are limited to a single frequency and polarization. To increase the signal-to-noise ratio and reduce the size of the data array, we average together 500 adjacent $16~\mu \text{s}$ samples, resulting in an effective time resolution of $8~\text{ms}$.

During drone flights, the calibration signal from the drone is chopped approximately 10 times per second ($T\approx10~\text{Hz}$), such that we can background subtract the telescope data acquired during the flight to isolate the contribution to the total power that comes from the calibration signal. Using a Pearson R (PR)\footnote{\href{https://docs.scipy.org/doc/scipy/reference/generated/scipy.stats.pearsonr.html}{Pearson R}} fitting routine, we find a solution for the best-fit square wave that best describes the pulsing behavior. This enables us to categorize each data point into one of three discrete cases. The first case includes integration periods where the calibration signal is always off. These data points represent measurements of the background signal power and are used to construct a time-varying background array. The second case includes integration periods where the calibration signal from the drone payload is always on. After subtracting the time varying background power from this subset of samples, we obtain the background subtracted measured power as a function of time. The final case includes times when the calibration signal is turning on or off during the integration period. These data points are not suitable measurements of the calibration signal or the background, and they are excluded from all future analysis.

\subsection{Combining and Synchronizing \alb and \drone data}

In order to produce a map, we are now tasked with combining the \alb and \drone data, each of which contains a GPS-derived time axis. In theory, these two data streams should be a priori synchronized, but in practice, it is not unusual for there to be an overall offset that exists between them. We obtain the best-fit timing offset between the two time axes using another PR fitting routine. Specifically, we manually vary a time offset that is added to the drone timestamps, and find the offset that maximizes the Gaussianity of the main beam (as described in \cite{ToneBeamcals}). With the two data streams synchronized, we proceed to interpolate each variable in the drone sensor data at each timestamp in the \alb autocorrelation data, thereby associating each power measurement with a drone position. This interpolation is reasonable because the drone flies at a low ($v\leq5~\text{m/s}$) and consistent velocity during long transits over the antenna.

\begin{figure*}
\centerline{\includegraphics[width=\textwidth]{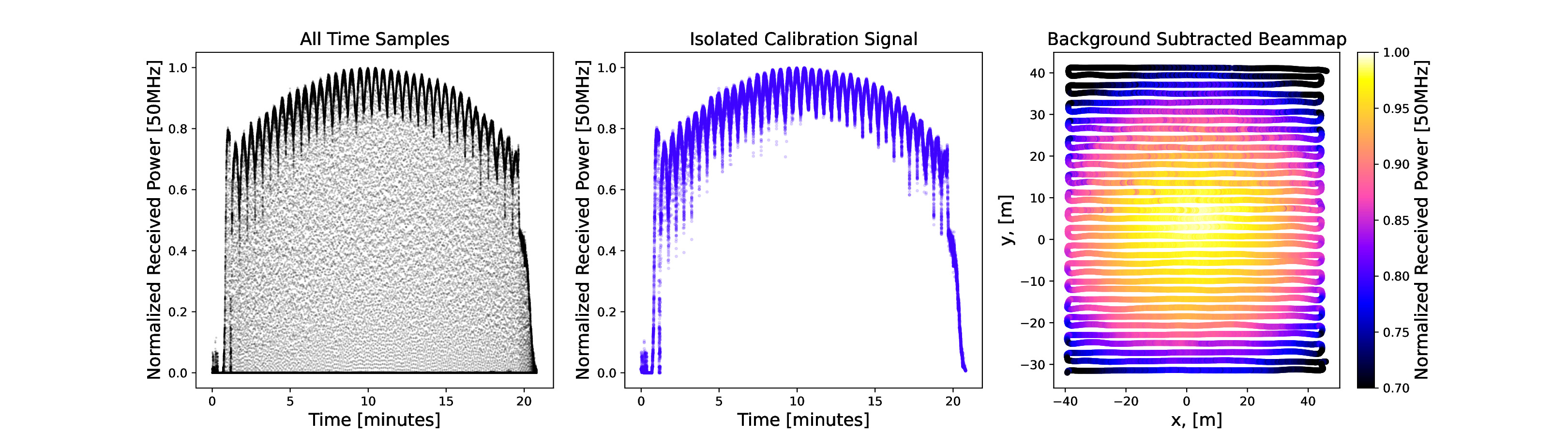}}
\caption{Raw data (left) from the 50~MHz channel of an \alb station acquired during a drone beam mapping flight at \mars during 2023. All time samples are shown in the raw data, including background measurements (values near zero) where the calibration source onboard the drone is off, measurements where the calibration source is on (high values) that follow the profile of the antenna beam, and transition measurements (intermediate values) where the source turns on or off during the integration time. By removing the transition and background measurements, we can isolate the calibration signal (center) in the time domain and associate each measurement with the drone's concurrent position to reveal the beam pattern of the antenna (right).}
\label{fig:timeseries}
\end{figure*}

During a demonstration measurement conducted at \mars in 2023, we measured the beam of the central \alb station's LWA antenna (see Figure~\ref{fig:timeseries}). To illustrate certain steps of data processing, compare the raw data (left, containing all time samples) to the isolated calibration signal (center, containing only ``source on'' time samples) which follows the Gaussian profile of the main beam. The rising and falling edges of the time series data correspond to the takeoff and landing of the drone, respectively, and are were not used to make beam maps. It is also notable that compared to the drone-mounted calibration source, the background signal level at \mars is negligible (approximately zero) throughout the acquisition. After synchronizing the drone and telescope data, and interpolating the drone data to the time resolution of \alb, we can produce a beam map (right) of the received power in the $50~\text{MHz}$ frequency channel for the N-S polarization of the \alb crossed dipole as a function of drone position. The transmitting antenna onboard the drone is linearly polarized and was aligned with the dipole axis during flight, yielding a measurement of the co-polar main beam, giving rise to the Gaussian structure present here. From this figure, we can directly assess the relative power of the calibration signal when measured in different portions of the antenna beam. 

\subsection{Removing Compression from Drone Measurements}

Several calibration steps must be applied to the background subtracted beam pattern to account for variations in the apparent intensity of the calibration signal, including distance compensation (accounting for geometric losses due to the changing distance between the drone and the antenna under test during the drone flight) and probe compensation (accounting for the known beam pattern of the transmitting antenna onboard the drone). When we attempted to compensate for this changing distance, the corrections applied at the edge of the beam were distorting the beam map significantly, causing power measurements from the edge of the beam to exceed power measurements obtained at boresight, contradicting the trend visible in the raw data. This implied that the calibration signal power was too powerful for the the front-end amplifiers of the \alb antenna, and were driving them into a non-linear gain regime. As a result, the measurements acquired during this drone flight were compressed, and in order to proceed with several well-understood calibration procedures we would need correct for the non-linearity of the amplifiers and the resulting compression.

\begin{figure}
    \centering
    \includegraphics[width=\textwidth]{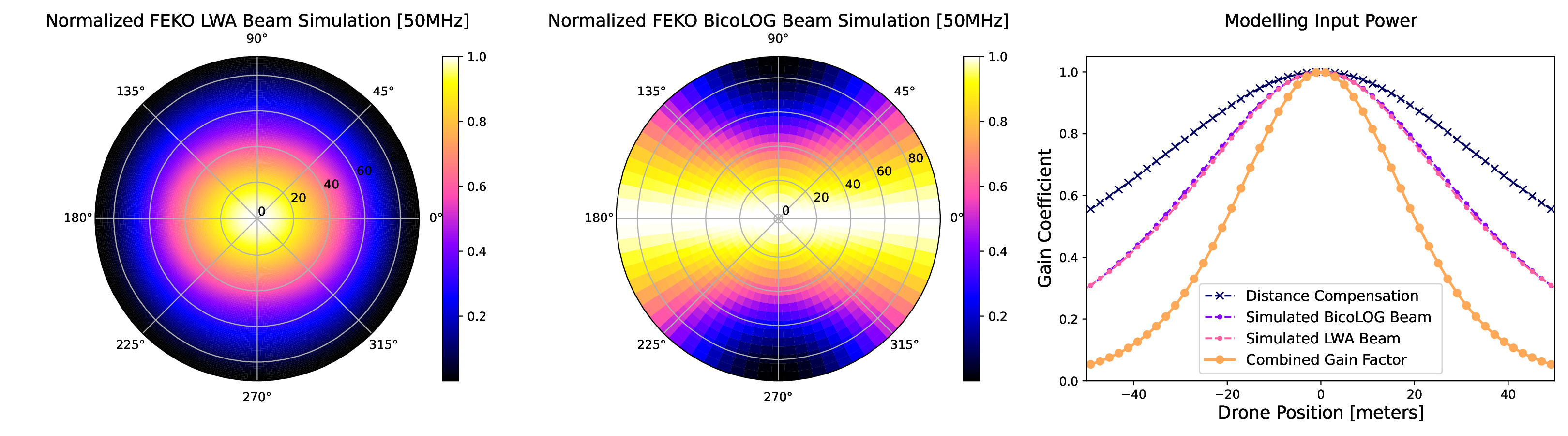}
    \caption{FEKO simulations of N-S dipole of the LWA antenna at 50~MHz (left) installed at each \alb station and the BicoLOG antenna (center) used to transmit the calibration signal from the \drone drone. The beam pattern of the LWA antenna is broad and Gaussian, while the BicoLog beam pattern can be reasonably approximated as a dipole that radiates nearly isotropically around its horizon but has diminished sensitivity approaching its poles. The spatial coverage of this drone flight (an 80~m x 80~m square at a constant altitude of 60~m) spanned approximately 33 degrees on either side of zenith, near the center of both beam simulations. The combined effect of these attenuation processes can be modelled (right) as a function of the drone's position, which changes as it progresses through its flightplan. We model the expected attenuation (gain) from the changing distance (blue), the BicoLOG beam (pink), and the LWA beam (purple), as well as their combined attenuation (yellow) during a transit across the zenith in the y-dimension. Examining the combined gain factor as a function of position, we should expect power measurements acquired near the edge of our beam map (-40~m or +40~m) to be around 10\% of the value measured at boresight (0~m) rather than the 75\% observed in our compressed data (see Figure \ref{fig:timeseries}).}
    \label{fig:fekobeams}
\end{figure}

To measure and remove the compression from our measured data, we began by modelling the input power associated with each power measurement. During the beam mapping flight, the calibration signal onboard the drone is broadcast at a constant signal level which--when measured by the \alb antenna--is attenuated by three independent processes: the beam of the receiving antenna, the beam of the transmitting antenna, and the changing distance between them. It is straightforward to estimate the attenuation from the changing distance between the drone and the \alb antenna using the GPS-derived drone location and scaling the incident power according to the inverse square law. Using FEKO simulations of the LWA and BicoLOG beams at 50~MHz we can estimate the attenuation that arises from the antenna beams by sampling the drone coordinates associated with each power measurement. The coordinates of the drone in the reference frame of the stationary \alb antenna are easily obtained from the GPS-derived drone location, whereas the location of the \alb antenna in the reference frame of the moving drone must account for drone's yaw, pitch, and roll. Both FEKO simulations and the modelled input power as a function of drone location are presented in Figure \ref{fig:fekobeams}).

\begin{figure}
    \centering
    \includegraphics[width=\textwidth]{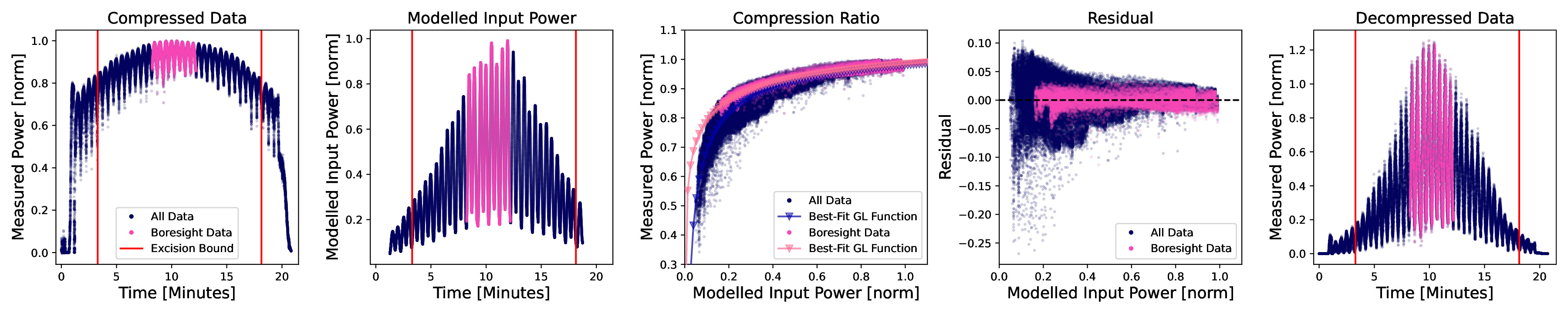}
    \caption{The procedure used to remove the nonlinear amplifier compression from the beam measurements. In each panel, the full data from the drone flight are shown in blue, while a subset of points near the zenith are shown in pink. Red vertical lines appear in each time domain plot, denoting time samples that contain the drone's takeoff and landing which are excluded from further analysis. Initially, the background subtracted power measured by the \alb antenna was compressed (left). FEKO beam simulations were sampled at the corresponding drone locations to estimate the input power to the antenna (second from left) as a function of time. Examining measured power as a function of input power, a non-linear compression relationship is observed (center), which has been fit with a Generalized Logistic Function. The corresponding residuals (second from right) are tight and unbiased for the subset of measurements acquired when the drone was near zenith, and have thus been used to calibrate the full flight data. The compressed data can be `decompressed' (right) by applying the inverse of the best-fit Generalized Logistic Function to each power measurement.}
    \label{fig:decompression}
\end{figure}

The decompression procedure applied to our beam data is illustrated in Figure \ref{fig:decompression}. For each timestamp, we model the input power incident on the \alb antenna by evaluating the FEKO beam simulations and compensating for the distance using the position and orientation of the drone. Examining the relationship between the measured power and the modelled input power, we can observe that a smooth, monotonic, and logarithmic (rather than linear) relationship exists between these two variables. To characterize the compression response of the amplifiers, we fit this plot with a Generalized Logistic Function that approaches a horizontal asymptote of one as the modelled input power increases. Within this figure, all data are shown in blue, while a selected subset of data points when the drone is near zenith are shown in pink. This subset was selected for characterizing the amplifier compression relationship because it is tightly clustered in this coordinate space and has low magnitude and unbiased residuals when fit with the generalized logistic function. Moreover, the time samples used in the final beam maps (see Section \ref{sec:results}) denoted by the red excision bounds have high measured power values, and are thus well characterized by this best-fit solution. To `decompress' the data, we apply the inverse of the best-fit Generalized Logistic Function to the `compressed' data shown in the first panel. 

Ultimately, the decompression procedure utilized in this analysis was only sensitive to the measured power (rather than any position or time dependent modeling) to avoid introducing bias. The drone position and orientation, the simulated antenna beams, and the distance compensation factors were not used directly in rescaling the raw data. While we have corrected this issue with caution and diligence, this analysis emphasizes the importance of selecting an appropriate amplitude for the calibration signal before measuring the instrument beam.

\subsection{Pixelization Scheme}
Understanding the structure of the instrument beam with high precision requires additional processing beyond the time domain representation shown in preliminary beam maps (see Figure \ref{fig:timeseries}). We can combine multiple time-domain baseband measurements acquired within the same portion of the beam and condense the statistical power of our measurements to improve the signal-to-noise. The result of this pixelization is an unambiguous relation for the beam amplitude as a function of position in the instrument field of view.

The choice of coordinates for the pixelization scheme is arbitrary, and for our purposes it is convenient to bin the data within a 2-dimensional Cartesian coordinate system with cells that are 2.5 square meters (approximately 2.4 degrees at zenith) in size. This follows the shape of the drone flight, and therefore preserves the number of baseband measurements (approximately 100 samples) that are averaged within a single pixel. 

Crucially, this processing step can combine data from multiple drone flights, which will be useful for two reasons in future measurement campaigns: evaluating the repeatability of the calibration signal across flights, and for co-adding data from multiple flights over the same dish---achieving higher signal-to-noise than is possible for a single drone flight. 

%%%%%%%%%%%%%%%%%%%%%%%%%%%%%%%%%%%%%%%%%%%%%%%%%%%%%%%
%%                      Results                      %%
%%%%%%%%%%%%%%%%%%%%%%%%%%%%%%%%%%%%%%%%%%%%%%%%%%%%%%%

\section{Results}
\label{sec:results}

\begin{figure}
    \centering
    \includegraphics[width=\textwidth]{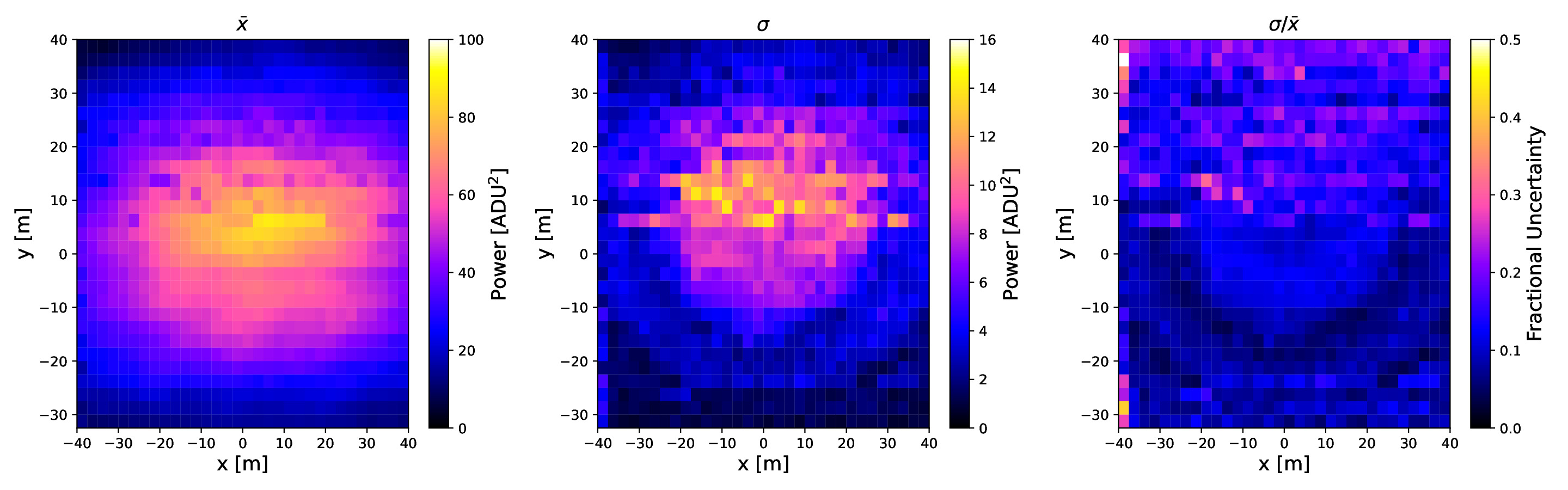}
    \caption{Pixelized beam map of the N-S polarization of an LWA antenna at 50~MHz. The field of view sampled during the drone flight has been divided into a Cartesian grid. Each 2.5 square meter pixel combines $\sim 100$ baseband measurements into a single averaged power value. The mean (left), standard deviation (middle), and their ratio (right) are shown for each pixel. Within the mean-normalized standard deviation map, pixels in the top half of the map have higher standard deviations, while pixels in the bottom half of the plot are consistently around the 10\% level. We attribute this inconsistency to a timing drift in the calibration signal across the duration of the drone flight. To rectify this issue in future drone flights we have modified the RF payload.}
    \label{fig:beam}
\end{figure}

The result of the data processing pipeline is a pixelated beam map where each pixel value is the average of all (decompressed and background-subtracted) baseband measurements acquired while the drone was within the pixel boundary. The pixelization scheme utilized in this analysis was a grid in Cartesian coordinates, comprised of  $2.5~\text{m}^2$ square meter pixels at a constant altitude of $60~\text{m}$. The beam map, shown in Figure \ref{fig:beam}, contains three panels: the mean (left), standard deviation (middle), and their ratio (right) are calculated per pixel. The data shown in this figure are from the $50~\text{MHz}$ frequency bin of N-S polarization channel, receiving power from a single linearly polarized transmitter onboard the \drone drone. The spatial extent of these plots ($-40\leq\text{x}\leq40$ and $-32.5\leq\text{y}\leq40$) has been cropped to eliminate empty pixels that were never measured by the drone. (The flight progressed from North to South, and was manually halted when the drone battery reached its safety margin of $20\%$ capacity.)

The mean beam map ($\bar{x}$) has recovered the broad Gaussian response of the LWA antenna at $50~\text{MHz}$, qualitatively matches our expectations from simulations (see Figure \ref{fig:fekobeams}). The standard deviation ($\sigma$) beam map closely follows the gaussian shape of the mean beam map, as expected, however some structure is present in the upper half of the plot ($5\leq\text{y}\leq40$), where the standard deviation exceeds what can be attributed to the Gaussian main beam. We attribute these high standard deviation values to the timing drift we detected in the calibration signal. When the calibration pulse period drifts, some power from the calibration signal leaks into our background measurements, resulting in an over-estimate of the (typically negligible) background signal level. When we subtract the background from neighboring calibration signal measurements, their variance increases (this effect is visible in the central panel of Figure \ref{fig:timeseries}, where the normalized received power is suppressed during the earlier portions ($3\leq\text{t}\leq10$) of the drone flight.) When the calibration pulse period was stable during the latter half of the drone flight ($-22.5\leq\text{y}\leq5$) the standard deviation values were substantially reduced.

Examining the mean-normalized standard deviation ($\sigma/\bar{x}$) beam map, we can produce a qualitative estimate of the repeatability of these measurements during a single drone flight. Within each pixel, we find a fractional uncertainty of approximately $10\%$ when the timing of the calibration pulses is stable. While this is higher than we hoped to achieve in these demonstration measurements, we are optimistic that we will reach a $5\%$ precision level in future drone measurements after implementing a few specific improvements to the timing and amplitude of the calibration signal (see Section \ref{sec:conclusions}). 

\begin{figure}
    \centering
    \includegraphics[width=\textwidth]{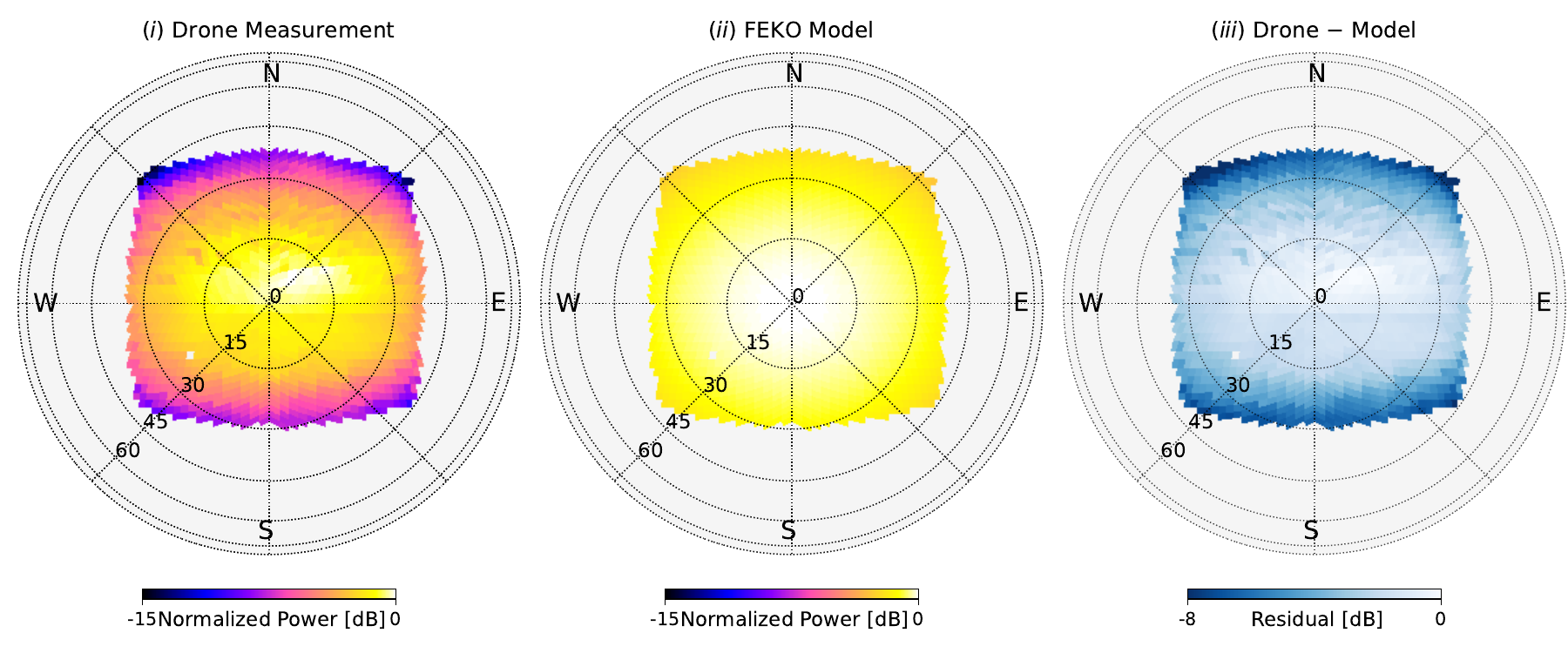}
    \caption{Comparison of the measured 50 MHz drone beam with the corresponding simulated FEKO model after projecting both onto a common HEALPix all-sky grid. Both beams are normalized to their peak response (0 dB). The residual map shows the normalized difference (Drone -- Model) over the measured footprint.}
    \label{fig:beam_model_comp}
\end{figure}

Finally, the calibrated beam map is projected onto a {\tt HEALPix}~\cite{2005ApJ...622..759G} all-sky grid with $\mathrm{NSIDE}=32$ (angular resolution $\approx110$ arcmin) to enable a direct pixel-by-pixel comparison with the simulated FEKO beam model at $50~\mathrm{MHz}$ over the measured footprint (Figure~\ref{fig:beam_model_comp}). Both the measured and simulated beams are normalized to their peak response prior to comparison. The measured beam reproduces the broad morphology of the simulated main beam, while the residuals become increasingly negative toward the edge of the mapped footprint, reaching approximately $-8~\mathrm{dB}$.

\section{Conclusions and Future Work}
\label{sec:conclusions}

In this publication we introduced the \drone drone system and demonstrated its capabilities by performing a beam calibration measurement in the high Arctic. The custom-built architecture has enabled on-site optimization, modifications, and repair work that would otherwise be difficult or impossible with commercial, off-the-shelf systems that are not designed to be serviced by the user. However, using a custom-built system also requires increased discipline in pre-flight checks (e.g., checking for connections that loosen with time from vibrations) and careful monitoring of the flight logs for any unexpected changes.

Despite several calibration challenges---most notably the calibration signal timing drift and the amplifier compression---we have reached a fractional uncertainty of approximately $10\%$ in our co-added beam map. Although these results are preliminary, they were produced with an \alb specific data processing and analysis pipeline, developed in preparation for future beam mapping campaigns. Now that the 8-element array is complete, we intend to characterize the beams of multiple \alb antennas in both polarizations across the primary science band ($4-30~\text{MHz}$) to a fractional uncertainty of $5\%$. We have identified several specific software and hardware improvements that we are pursuing to achieve this level of precision.

%%%%%%%%%%%%%%%%%%%%%%%%%%%%%%%%%%%%%%%%%%%%%%%%%%%%%%%
%%                    Future Work                    %%
%%%%%%%%%%%%%%%%%%%%%%%%%%%%%%%%%%%%%%%%%%%%%%%%%%%%%%%

%\section{Future Work}
%\label{sec:futurework}

In the near future, we plan to return to MARS to conduct a new beam mapping campaign, with several goals, including: measuring the beam of an \alb antenna in both polarizations out to a zenith angle of 80 degrees, assessing the repeatability of our drone measurements by performing multiple calibration flights over a single antenna, and comparing the beams of \alb antennas installed at different stations. In preparation, we have already made numerous improvements to the analysis pipeline and the \drone drone system. 

The data processing and analysis presented in this paper is ready to support future drone beam mapping campaigns. Because this software is already written, tested, and incorporated into an open source module, we will be able to produce beam maps within hours of conducting a drone flight. Armed with the capacity to inspect drone flight data while still deployed in the arctic, we will be promptly alerted to problems like the amplifier compression present in our existing flight data (discussed in Section \ref{sec:processing}). To avoid repeating this specific mistake in future meeasurements, we have measured the compression profile of our front-end electronics in the lab, and will appropriately attenuate our calibration signal. 

Previously, the pulsing cadence of the RF payload was governed by issuing sleep commands (via GPIO) to an RF switch (as described in Figure \ref{fig:payload}). This method of chopping was inconsistent, resulting in the pulse cadence drifting over the course of a drone flight. To remedy this issue, we have developed a new payload, comprised of an Arduino and an Si5351 clock generator chip\footnote{\url{https://learn.adafruit.com/adafruit-si5351-clock-generator-breakout/overview}} that can generate an RF calibration signal across a wide frequency range (8~KHz to 150~MHz). The output of the clock generator is now pulsed in software, synchronized to the Arduino's 16~MHz crystal oscillator, which will accrue a negligible drift over the same timescale. This improvement in timing precision will translate to improved precision in identifying calibration pulses in the baseband data, more accurate background subtraction, and higher signal-to-noise in our final beam maps. Additionally, this payload can be remotely control, which would enable us to change the properties of the calibration signal without landing the drone.

A final avenue we are pursuing is extending the use case for the \drone platform beyond the \alb experiment. The methodology and techniques presented here for drone-based beam calibration are applicable across a wide range of the radio spectrum, with a key difference being the frequency range of the calibration signal. We have produced additional copies of the \drone system with bespoke RF payloads for calibrating two other $21~\text{cm}$ instruments that operate at much higher frequency ranges: CHORD\cite{CHORD_Instrument} ($300~\text{MHz}$ to $1500~\text{MHz}$) and HIRAX\cite{hirax} ($400~\text{MHz}$ to $800~\text{MHz}$). We also plan to extend these methods to much lower frequencies ($\sim4-30~\text{MHz}$) to calibrate the primary science band of \alb using a new transmitting antenna. We are still investigating commercially available antennas that have favorable properties at these frequencies while remaining compact, lightweight, and drone-mountable.

\section*{ACKNOWLEDGMENTS}
\label{sec:acknowledgements}

We gratefully acknowledge the Natural Sciences and Engineering
Research Council of Canada (RGPIN-2019-04506, RGPNS-2019-534549), the
Canada Foundation for Innovation John R.~Evans Leaders Fund (40824),
the New Frontiers in Research Fund (NFRFE-2021-00409), the National
Geographic Society (NGS-949383T-22), and the Polar Continental Shelf
Program (PCSP) for providing funding and logistical support for our
research program.  This research was undertaken, in part, thanks to
funding from the Canada 150 Program.  This research was enabled in
part by support provided by the Digital Research Alliance of Canada
(alliancecan.ca).  We extend our sincere gratitude to the PCSP
Resolute staff for their generous assistance and bottomless cookie
jars.  The authors thank Chris Omelon, Laura Thomson, Anthony Zerafa,
Olivia Locke, and all of the \mars\ researchers for their invaluable
advice and field help.

%%%%%%%%%%%%%%%%%%%%%%%%%%%%%%%%%%%%%%%%%%%%%%%%%%%%%%%
%%                 A P P E N D I X                   %%
%%%%%%%%%%%%%%%%%%%%%%%%%%%%%%%%%%%%%%%%%%%%%%%%%%%%%%%

\section{Appendix: ionosphere error}\label{subsec:ionosphere_appendix}
Consider an interferometric observation of a bright, unresolved source with flux density $F$ using two antennas $i$ and $j$, and assume that the source dominates the antenna beams. If the source is in the phase center of the interferometric baseline, the correlated power measured by two antennas, referred to as visibility $V$, can be written as
\begin{equation}
    V = g_i g_j^* E_i E_j^*
\end{equation}
where $g_i$, $g_j$ are complex electric field beams of the antennas in the direction of the source in units of meter\footnote{Effective area of the antenna in a given direction is proportional to $g^2$.} and $E_i$, $E_j$ are the incident electric fields. The phase of one of the $E$ fields is assumed to have been rotated to account for geometrical path length difference needed to bring the source to the phase center. In this work, the power beam of the antenna was characterized, which provides only the amplitude of the gains $g$. To compare the effect of ionospheric errors and power beam measurement errors, and develop a criterion for an acceptable level of beam errors, we conduct a purely amplitude-based error analysis.

In what follows, let $g_i$ and $g_j$ be real-valued quantities. The amplitude of the measured visibility at any given time can be written as:
\begin{equation}
    |V| = g_i g_j \sqrt{F} (1 + a_i) \sqrt{F} (1 + a_j)
\end{equation}
where $a_i$ and $a_j$ encode the fractional error on incident electric field amplitude due to ionospheric scintillation. If there is little to no absorption in the ionosphere, which is a good assumption during quiet polar night times, then the power received by any given antenna on the ground must be conserved. Or
\begin{equation}
    \langle (1 + a_i)^2 \rangle = \langle (1 + a_j)^2 \rangle = 1.
    \label{eq:power_cons}
\end{equation}
This can be understood by considering waves reaching the ground from the ionospheric can be thought of as a linear combination of many stochastic diffraction patterns. A point on the ground is some times in a bright spot, and some other times in a dark spot. Over the entire ground, however, the total flux must be the same as the source flux since scattering can only re-distribute energy\footnote{This is strictly true only for single antenna measurements. An interferometer can resolve out parts of the scatter-broadended source. Since phases are ignored in this analysis, we limit our attention to single antenna averages.}. As plasma density structures causing the scattering drift over the field-of-view of the antennas, by the principle of ergodicity, ensemble average of source flux at one spot formed by means of time-averages must also approach the same value. See discussion surrounding equations 3.14 and 3.43 in \cite{yeh1982radio}.

Using \eqref{eq:power_cons}, we have for the mean value of the detected amplitude in the presence of scintillation:
\begin{equation}
    \langle a_i \rangle = -\frac{1}{2} \langle a_i^2 \rangle,
    \label{eq:mean_scint}
\end{equation}
and similarly for $a_j$. This tells us that incoherent averaging of electric field amplitudes in the presence of scintillation will converge to a value less than its true value, which is often empirically observed \cite{whitney1972estimation}. Using the measured visibility, and measured gains, we'd like to form an \textit{estimate} of the source flux, $\Hat{F}$. The drone-based measurement technique demonstrated here measures the power beam. Hence the actual measurement error in the power domain, i.e. $\hat{G}_i = g_i^2 (1 - \delta_i)$, where $\delta_i$ is the random fractional error in power beam measurement. An estimate of the amplitude beam can be formed by taking the square-root of the power beam. Let the estimate of amplitude gains be $\Hat{g}_i$ and $\Hat{g}_j$, related to true gains as $\Hat{g}_i = g_i \sqrt{1 - \delta_i} $ and $\Hat{g}_j = g_j \sqrt{1 - \delta_j} $. Then to first order in the $\delta$'s, we have:
\begin{equation}
    \Hat{F} = \frac{|V|}{\Hat{g}_i \Hat{g}_j} = F\frac{(1 + a_i)(1 + a_j)}{\sqrt{1 - \delta_i}\sqrt{1 - \delta_j}} \approx F \underbrace{(1 + a_i)(1 + a_j)}_{X}\underbrace{(1 + \delta_i/2) (1 + \delta_j/2)}_{Y}.
\end{equation}
$X$ denotes the error contribution due to ionospheric scintillation and $Y$ denotes that due to beam measurement errors. Since the two are independent, the task of evaluating variance of $\Hat{F}/{F}$ is reduced to evaluating individual variances $V_X$ and $V_Y$, such that the net variance is
\begin{equation}
    \mathrm{Var}\frac{\Hat{F}}{F} = V_X \mu_Y^2 + V_Y \mu_X^2 + V_X V_Y.\label{eq:full_var}
\end{equation}
where $\mu$'s denotes the mean $\langle . \rangle$ of respective quantities. 

\subsection{Evaluating Scintillation Term}
The two terms in $X$ can be correlated if the two antennas are seeing through the same patch of ionospheric disturbance, for calculating $V_X$, we use the full covariance relation \cite{bohrnstedt1969exact}. The exact form of $V_X$ includes third and fourth order moments of the fluctuations $\Delta a_i$ and $\Delta a_j$ in $a_i$ and $a_j$ respectively. We make the assumption that the fluctuations are small, such that these higher moments can be ignored. This is applicable in the weak-scattering regime where $\mathrm{S}_4 \ll 0.5$. For notational brevity reasons, let $\alpha = \langle a_i^2 \rangle$. Variance $V_X$, to first order in variances, is given by
% \begin{equation}
%     V_X = 2 \alpha (1 - \alpha/2)^2 (1 - \alpha/4) (1 + \rho) \label{eq:var_x}
% \end{equation}
\begin{equation}
    V_X = 2 \alpha (1 + \rho) + \mathcal{O}(\alpha^2), \label{eq:var_x}
\end{equation}
where $\rho$ is the normalized correlation of the ionospheric scintillation at two antennas, defined as:
\begin{equation}
    \rho = \frac{\mathrm{Cov}(a_i, a_j)}{\sqrt{\mathrm{Var}(a_i)\mathrm{Var}(a_j)}}.
\end{equation}
Similarly, we can calculate the mean as
\begin{equation}
    \mu_X = 1 -\alpha(1 - \rho) + \mathcal{O}(\alpha^2).\label{eq:mu_x}
\end{equation}

\subsection{Converting Amplitude Errors to $\mathrm{S}_4$}
The scintillation index $\mathrm{S}_4$ at an antenna $i$ is defined as
\begin{equation}
    \mathrm{S}_4^2 = \frac{\mathrm{Var}(|E_i|^2)}{\langle |E_i|^2 \rangle^2} = \frac{\mathrm{Var}(1 + a_i)^2}{\langle (1 + a_i)^2 \rangle^2}.
\end{equation}
Using \eqref{eq:power_cons}, the above equation simplifies to
\begin{equation}
    \mathrm{S}_4^2 = \langle (1 + a_i)^4 \rangle - 1,
\end{equation}
which to first order in $a_i^2$ lets us write
\begin{equation}
    \mathrm{S}_4 = 2 \sqrt{\alpha}. \label{eq:s4_beam}
\end{equation}
We assume that prevailing ionospheric conditions above the two antennas are similar, such that a single value of $\mathrm{S}_4$ can be used to characterize the strength of scintillation. This does not preclude the possibility of the exact realization of scintillation noise in the two antennas being uncorrelated, i.e. $\rho \sim 0$, if they are very far apart from each other and see through uncorrelated patches of the ionosphere. 

\subsection{Evaluating Beam Term}
If $\delta_i$'s are treated as approximately zero-mean, then we have for the mean and variance of the beam term $Y$
\begin{equation}
    \mu_Y = 1,\label{eq:mu_y}
\end{equation}
and
\begin{equation}
    \mathrm{Var}(Y) = \frac{\sigma_\delta^2}{2} + \frac{\sigma_\delta^4}{16},\label{eq:var_y}
\end{equation}
where $\sigma_\delta^2$ is the variance of $\delta_i$.

Finally, substituting \eqref{eq:mu_x}, \eqref{eq:var_x}, \eqref{eq:s4_beam}, \eqref{eq:mu_y}, \eqref{eq:var_y} in \eqref{eq:full_var}, and retaining only the terms first-order in variance, we have that the fractional error in measured source flux is
\begin{equation}
    \frac{\sqrt{\mathrm{Var}(\Hat{F})}}{F} = \sqrt{\frac{\mathrm{S}_4^2 (1+\rho) + \sigma_\delta^2}{2}}.\label{eq:src_flux_error}
\end{equation}
For the initial beam measurements, we therefore target $\sigma_\delta \ll \mathrm{S}_4$ as a quantitative starting point to our desired precision. Note, however, that the relation given by \eqref{eq:src_flux_error} does not represent a fundamental limit to interferometric imaging precision. In the weak scintillation regime, complex gain calibration solutions can be recovered provided sufficient SNR is achieved within the coherence time set by ionospheric amplitude and phase fluctuations. Better understanding of the effects of polar ionosphere on observations of bright sources with ALBATROS will inform future revisions to this metric.

% to avoid being limited by beam uncertainties in amplitude measurements. 

% In the weak scintillation regime, complex gain calibration solutions can be obtained for observations of a source if high enough SNR can be achieved within the coherence time limitation imposed by amplitude and phase variability.

\bibliographystyle{ws-jai}
\bibliography{sample}{}

\end{document}